\documentclass[12pt]{article}
\usepackage{amssymb}
\usepackage[dvips]{epsfig}
\newcommand{\be}{\begin{equation}}
\newcommand{\ee}{\end{equation}}
\newcommand{\bn}{\begin{eqnarray}}
\newcommand{\en}{\end{eqnarray}}

\newcommand{\xt}{\tilde{x}}

\newcommand{\nn}{\nonumber}

\newcommand{\no}{\noindent}
\newcommand{\ylz}{YLZ }
\newcommand{\yles}{YLES }

\def\bea{\begin{eqnarray}}
\def\eea{\end{eqnarray}}
\def\lbd{\lambda}
\def\sg{\sigma}
\begin{document}

\title{\textbf{Critical behavior at edge singularities in one dimensional spin models}}
\author{D. Dalmazi\footnote{dalmazi@feg.unesp.br}  and F. L. S\'a\footnote{ferlopessa@yahoo.com.br} \\
\textit{UNESP - Campus de Guaratinguet\'a - DFQ } \\
\textit{Av. Dr. Ariberto P. da Cunha, 333} \\
\textit{CEP 12516-410 - Guaratinguet\'a - SP - Brazil. }}
\maketitle

%\footnote{One exception is the Q-state Potts model, see
%\cite{uzelac,sykim}, where $Q-2$ transfer matrix eigenvalues are
%degenerate and independent of the magnetic field. The other two
%eigenvalues which depend on $H$ are roots of a simple second
%degree polynomial which makes this case fairly simple and similar
%to the $S=1/2$ Ising model.}
%\bibitem{sykim} S-Y. Kim, Journal of the Korean Phys. Society,
%{\bf 45} (2004) 302-309, see also cond-mat/0305456.

\begin{abstract}

In ferromagnetic spin models above the critical temperature ($T >
T_{cr}$) the partition function zeros accumulate at complex values
of the magnetic field ($H_E$) with  a universal behavior for the
density of zeros $\rho (H) \sim \vert H - H_E \vert^{\sg}$. The
critical exponent $\sg$ is believed to be universal at each space
dimension and it is related to the magnetic scaling exponent $y_h$
via $\sg = (d-y_h)/y_h$. In two dimensions we have $y_h=12/5 \,
(\sg = -1/6)$ while $y_h=2 \, (\sg=-1/2)$ in $d=1$. For the one
dimensional Blume-Capel and Blume-Emery-Griffiths models we show
here, for different temperatures, that a new value  $y_h=3 \, (\sg
=-2/3)$ can emerge if we have a triple degeneracy of the transfer
matrix eigenvalues.

\end{abstract}

\vfill\eject

\section{Introduction}

In \cite{yl} C. N. Yang and T.D. Lee have initiated a method to study phase transitions based on partition
function zeros. Since then, several applications have appeared in equilibrium and more recently in
non-equilibrium  statistical mechanics, see \cite{bena} for a review and references. It is expected that with
the increase of computer power new applications will appear specially where analytic methods can hardly be used.
In many statistical models the so called Yang-Lee zeros (\ylz) lie on arcs on the complex u-plane,
$u=\exp(-\beta H)$, where $H$ is an external magnetic field and $\beta=1/K_B T$. The zeros at endpoints of the
arcs tend to pinch the positive real axis $\Re(u)> 0$ at the critical value of the magnetic field for $T \le
T_{cr}$ while for $T > T_{cr}$ they accumulate at complex values $u_E=\exp (-\beta H_E)$ which are called
Yang-Lee edge singularities (\yles). The linear density of zeros has a universal power like divergence $\rho (u)
\sim \vert u - u_E \vert^{\sg}$ in the neighborhood of the \yles. The critical properties of the \yles are
described by a non-unitary $\imath \varphi^3$ field theory \cite{fisher} which corresponds in two dimensions
\cite{cardy} to a $(p,q)=(2,1)$ non-unitary conformal minimal model leading to the prediction $\sigma=-1/6$.
This is in agreement with numerical results for the ferromagnetic Ising model, see e.g. \cite{ms,kim2006}, and
experimental magnetization data \cite{binek}. Even in the anti-ferromagnetic Ising model where $ u_E \in
\mathbb{R}_- $ numerical results \cite{kim2005} also lead to $\sg=-1/6$. Analogously, for the ferromagnetic and
anti-ferromagnetic $S=1/2$ Ising model in $d=1$ one derives exactly $\sg=-1/2$. In the ferromagnetic case, the
value $\sg=-1/2$ has been found in other $d=1$ models like the n-vector chain \cite{kurze} and the q-state Potts
model with $q > 1 $ and $ 0 \le q < 1$ \cite{uzelac}. Furthermore, in \cite{wang}, by assuming that the transfer
matrix eigenvalues are real in the gap of zeros and that the magnetic field is pure imaginary, one obtains
$\sg=-1/2$ for the $ S=1,3/2 $ Ising model and also for the Blume-Emery-Griffiths (BEG) model investigated here.
The same assumptions of \cite{wang} have been made in the proof of $\sg=-1/2$ given in \cite{fisher2,kurze}.
More recently \cite{ananikian} another proof which holds without those assumptions has appeared in the
Blume-Capel model. We show here that the latter proof contains also some hypotheses  which are not satisfied in
part of the parameter space of the BEG model where a new critical behavior appears.

\section{General setup and analytic results}

Our starting point is the couple of finite size scaling (FSS)
relations, see \cite{pearson,creswick}, :

\bea u_1(L) -  u_1 (\infty)\,   &=&  \, \frac{C_1}{L^{y_h}} \, +
\,\cdots   \label{un} \\
\rho (L) \, &=& \, C_2 \, L^{y_h-d} \, + \, \cdots  \label{rhon}
\eea

\no where $C_1 \, , \, C_2$ are constants, independent of the size
of the lattice $L$. The Yang-Lee zero $u_1(L)$ is the closest zero
to the edge singularity $u_E = u_1(\infty)$. The density of zeros
$\rho(L)$ is calculated at $u=u_1 (L)$. In practice we use

\be \rho (L) \, = \, \frac 1{N\vert u_1(L) - u_2(L) \vert }
\label{rho} \qquad , \ee

\no where $u_2(L)$ is the second closest zero to $u_E$ and $N$ is the total number of spins ($N=L^d$). The dots
in ($\ref{un}$),($\ref{rhon}$) stand for corrections to scaling. Our analytic method consists of deriving $y_h$
from an expansion for large number of spins similar to ($\ref{un}$) obtained directly from the transfer matrix
solution. We also calculate the YLZ numerically for finite number of spins and check the validity of
($\ref{un}$),($\ref{rhon}$) which furnish numerical estimates for $y_h$.

The partition function for the BEG model is given by \cite{beg}

\be Z_N\,=\,\sum_{\left\{S_i\right\}}\, \exp \beta\left\lbrace
J\sum_{\left<ij\right>}\, S_i\,S_j\,+\, K\sum_{\left<ij\right>}\,
S^2_i\,S^2_j\, + \, \sum_{i=1}^N\, \left\lbrack
H\,S_i\,+\,\Delta\,(1\,-\, S_i^2) \right\rbrack \right\rbrace
\quad , \label{zbeg}\ee

\no where $S_i=0,\pm 1$ and the sum $\sum_{\left<ij\right>}$ is
over nearest-neighbor sites. We assume ferromagnetic Ising
coupling $J>0$ but $K \, , \,  \Delta $ can have any sign. All the
couplings $J \, , \,  K \, , \, \Delta $  are real while the
magnetic field $H$ may be complex. The temperature is defined with
respect to the ferromagnetic coupling via the compact parameter
$c\equiv \exp (-\beta J)$, where $0 \le T < \infty $ corresponds
to $0 \le c \le 1$. In $d=1$, using periodic boundary conditions,
$S_i=S_{i+N}$, the model can be easily solved via transfer matrix
:

\be \, Z_N \, = \, c^{-N} \left\lbrack \lbd_1^N + \lbd_2^N +
\lbd_3^N \right\rbrack \quad , \label{zn} \ee

\no where the eigenvalues $\lbd_1 \, , \, \lbd_2 \, , \, \lbd_3 $
are the solutions of the cubic equation

\be \lbd^3 - a_2 \lbd^2 + a_1 \lbd - a_0 \, = \, 0 \qquad ,
\label{cubic} \ee

\no with coefficients

 \bea a_0 \, &=& \, b \, \xt (1-c^2)\left\lbrack b (1 + c^2)- 2
c\right\rbrack \nn\\
 a_1 \, &=& \, b^2 (1-c^4) + A \,\xt (b-c) \label{a2} \\
 a_2 \, &=& \, \xt + A \, b \nn \eea

\no We have defined for convenience

 \be b = \exp(\beta K) \quad , \quad x = \exp(\beta\Delta) \quad ,
 \quad \xt = x c =\exp\beta (\Delta - J) \label{ctes1} \ee

 \be A = u + 1/u = 2\cosh (\beta H) \label{A} \ee

\no The quantity $u=\exp(-\beta H)$ plays the role of fugacity in
a lattice gas \cite{yl}. The partition function $Z_N$ is
proportional to a polynomial of degree $2N$ in the fugacity.
Therefore, all relevant information about $Z_N$ is contained in
its zeros $Z_N(u_k)=0 \, , \, k=1,\cdots , 2N$. Due to
$\mathbb{Z}_2$ symmetry, $Z_N(u)=Z_N(1/u)$, half of the zeros are
the inverse of the other half. The exact position of the YLZ can
hardly be found even in one dimension. In the thermodynamic limit
the partition function will vanish whenever two or more
eigenvalues share the largest absolute value, see
\cite{nielsen,katsura,uzelac}. For large finite number of spins
the same conditions are still useful, see e.g. \cite{almeida}, for
locating the YLZ. Thus, following \cite{ananikian}, let us assume
that there exists a function $A(\varphi)=u(\varphi) +
u^{-1}(\varphi)$ such that when we plug it back in (\ref{cubic})
we have:

\be \lbd_2 \, = \, e^{i \, \varphi} \lbd_1 \label{5a} \ee \be
\vert \lbd_2 \vert \, = \, \vert \lbd_1 \vert \, > \vert \lbd_3
\vert \label{5b} \ee

\no Therefore, for a large number of spins we can neglect the
contribution of $\lbd_3$ and write $ Z_N \approx \vert \lbd_1
\vert^N \left( 1 + e^{i\, N\, \varphi} \right)$. The $2N$ YLZ are
found from

\be u^{\pm}(\varphi_k) \, = \frac 12 \left\lbrack A(\varphi_k) \,
\pm \, \sqrt{A(\varphi_k)^2 - 4} \right\rbrack \qquad , \label{6}
\ee

\no with $\varphi_k \, = (2k-1)\frac{\pi}N \, , \, k=1,2,\cdots ,
N$. In order to find $A(\varphi)$ we start from the  relations
$a_0=\lbd_1\lbd_2\lbd_3 \, $ , $\, a_1 = \lbd_1\lbd_2 +
\lbd_1\lbd_2 + \lbd_2\lbd_3 \, $ , $\, a_2 = \lbd_1 + \lbd_2 +
\lbd_3 $. Implementing the condition (\ref{5a}) and eliminating
$\lbd_3$ we obtain:

\bea a_1 \, &=& \, \lbd_1 a_2 (1 + e^{i \varphi}) \, - \, \lbd_1^2
\left( 1 + e^{i \varphi} + e^{i 2 \varphi}\right) \label{11a} \\
a_0 \, &=& \, \lbd_1^2 a_2 e^{i \varphi} \, - \, \lbd_1^3
\left(e^{i \varphi} + e^{i 2 \varphi}\right) \label{11b} \eea

\no Manipulating (\ref{11a}) and (\ref{11b}) we further eliminate
$\lbd_1$ and find an equation for $A(\varphi)$:

\be a_0^2 \left(1+2\cos\varphi \right)^3 + 4\cos^2\frac{\varphi}2
\left(a_1^3 + a_0 a_2^3 \right) - a_1^2 a_2^2 - 2 \left(1 + 2 \cos
\varphi \right)\left(2+ \cos \varphi \right)a_0 a_1 a_2 = 0
\label{fphi} \ee

\no which is the same expression obtained in \cite{ananikian} for
the Blume-Capel model ($b=1$). At this point, in principle, the
Yang-Lee zeros  are determined by (\ref{fphi}), once we check
(\ref{5b}). After taking the continuum limit in the corresponding
solution $u(\varphi_k)$ one could derive the density of zeros and
obtain the exponent $\sigma$. However, in practice, the expression
(\ref{fphi}) is a fourth degree polynomial equation for
$A(\varphi)$ whose solutions are cumbersome in general. It is not
feasible to substitute them in the solutions of the cubic equation
(\ref{cubic}), which are a bit complicated too, and then check the
condition (\ref{5b}) for arbitrary values of the parameters $b , c
, x$.

Fortunately, in order to compute the exponent $\sigma$ or $y_h$ we
only need to study the vicinity of the YLES. The YLES is located
at $A(\varphi=0)$ since at $\varphi=0$ the two, presumably,
largest eigenvalues of the transfer matrix coincide
($\lbd_1=\lbd_2$) which guarantees, see the original works
\cite{lassettre},\cite{lamb} and \cite{wang} for a recent work,
the existence of phase transition. Therefore, only the behavior of
the solutions of the equation (\ref{fphi}) about $\varphi =0 $ is
required, as already noticed in \cite{ananikian}, remembering of
course that (\ref{5b}) must hold. Nevertheless, differently from
those authors we do not take the continuum limit. By keeping $N$
finite we will be able to study the large $N$ expansion for the
closest zero to the YLES and determine the exponent $y_h$ via
comparison with (\ref{un}).

In order to scan the parameter space of the BEG model for a
possible new critical behavior at edge singularities we have found
more useful instead of using (\ref{fphi}) to take consecutive
derivatives of expressions (\ref{11a}), (\ref{11b}) and make some
mild hypotheses as shown next.

First of all, from the fact that the YLES is located at
$A(\varphi=0)$ it is natural to assume that the smallest phase
among all $\varphi_k$, i.e., $\varphi_1=\pi/N$, corresponds to the
closest zero to the YLES, namely $u_1(N) \, = \, u (\varphi_1)$.
The zero $u(\pi/N)$ is determined from $A(\pi/N)$ via (\ref{6}).
Thus, all we need is a large $N$ expansion for $A(\pi/N)$. Our
basic hypothesis is the existence of a region in the  parameter
space of the BEG model  where we are allowed to expand
$A(\varphi_1)$ in a Taylor\footnote{Notice that due to the square
root in (\ref{6}), when the YLES is located at $A(0)=-2$ ($u=-1$),
the existence of (\ref{expansion}) does not always lead to a
Taylor expansion for $u_1(N)$, as assumed in \cite{ananikian}}
series about $\varphi_1 = 0$:

\be A(\varphi_1) \, = \, A(0) + \varphi_1 \, \frac{d A}{d
\varphi}\vert_{\varphi =0} \, + \, \frac{\varphi_1^2}{2 !}
\frac{d^2 A}{d \varphi^2}\vert_{\varphi =0} \, + \, \cdots
\label{expansion} \ee

\no In order to proceed further we would like to collect
information about $d^n A/d\varphi^n $ at $\varphi=0$. From the
first derivatives of (\ref{11a}) and (\ref{11b}) we deduce:

\be \left\lbrack \xt ( b - c) - b\, \lbd_1 \right\rbrack \,
\frac{d A}{d \varphi} \, = \, \left\lbrack e^{\imath
\varphi}-1\right\rbrack
\frac{\partial\lbd_1}{\partial\varphi}\left\lbrack (2+
e^{\imath\varphi})\lbd_1 - a_2 \right\rbrack \label{14a} \ee

\no Assuming that $\partial\lbd_1/\partial\varphi$ at $\varphi=0$
is smooth enough to guarantee that the right handed side of
(\ref{14a}) vanishes at $\varphi=0$ we get:

\be \left\lbrack \xt ( b - c) - b\, \lbd_1
\right\rbrack_{\varphi=0} \, \frac{d A}{d
\varphi}\vert_{\varphi=0} \, = 0 \label{14b} \ee

\no Therefore if $\lbd_1 \ne \xt (b-c)/b $ we derive $\left(d
A/d\varphi\right)_{\varphi=0}=0$. Since at $\varphi=0$ we have
double degeneracy $\lbd_1=\lbd_2$, back in the cubic equation
(\ref{cubic}) it is possible to show that $\lbd_1=\lbd_2=\xt
(b-c)/b$ requires $\xt=(1-c^2)b^2/\vert b - c \vert $. Thus, if
$\xt \ne (1-c^2)b^2/\vert b - c \vert $ we have\footnote{It is
clear from the cubic equation (\ref{cubic}) that in the special
case $b=c$  the condition $\lbd_1 \ne \xt (b-c)/b $ is always
satisfied for any finite non vanishing temperature.} $\left(d
A/d\varphi\right)_{\varphi=0}=0$. Consequently, the
 second derivatives of (\ref{11a}) and (\ref{11b}) furnish

\be \left\lbrack \xt ( b - c) - b\, \lbd_1
\right\rbrack_{\varphi=0} \, \frac{d^2 A}{d
\varphi^2}\vert_{\varphi=0} \, = \, \left\lbrack \frac{\lbd_1}2
(\lbd_1 - \lbd_3)\right\rbrack_{\varphi=0} \label{14} \ee

\no Then, if there is no triple degeneracy $\lbd_1=\lbd_2 \ne
\lbd_3$ and $\xt \ne (1-c^2)b^2/\vert b - c \vert $ we end up with
$\left(d A/d\varphi\right)_{\varphi=0}=0$ but $\left(d^2
A/d\varphi^2\right)_{\varphi=0} \ne 0 $. Back in (\ref{expansion})
and then in (\ref{6})
 we derive  by comparison with
(\ref{un}) the usual result $y_h = 2$ and $\sigma =-1/2$ which
generalizes for the BEG model the proof of $\sigma =-1/2$ given in
\cite{ananikian} for the Blume-Capel model. On the other hand, if
the couplings are such that we have triple degeneracy, using
$\left(d A/d\varphi\right)_{\varphi=0}=0=\left(d^2
A/d\varphi^2\right)_{\varphi=0}$ it is easy to show that
$\left(d^3 A/d \varphi^3\right)_{\varphi=0} \ne 0 $, consequently
we have a new critical behavior with $y_h=3$ and $\sigma = -2/3 $
at the edge singularity. For completeness we mention the third
case $\xt = (1-c^2)b^2/\vert b - c \vert $ which leads to $\left(d
A/d\varphi\right)_{\varphi=0} \ne 0$ and $y_h=1$, consequently
$y_h -d = 0$.

Regarding the exponent $y_h$, the point $u_E=-1 (A(0)=-2)$, which
corresponds to $\beta H_E = \pm \imath \pi $, is special as we see
from (\ref{6}). Due to the square root in (\ref{6}) we expect a
change from $y_h$ to $y_h/2$ when the edge is located at $u_E=-1$.
The values $y_h=1,2,3$ will be replaced by $y_h=1/2,1,3/2$
respectively. Indeed, all such special values for $y_h$ at
$u_E=-1$  have been confirmed numerically, though they are not
presented here with the exception of $y_h=3/2$ which appears in
table 1 (third line from the bottom) for which $\sigma =-1/3$
instead of $-2/3$. Even in the cases where $y_h=1,1/2$ our
numerical results based on the FSS relations (\ref{un}) and
(\ref{rhon}), seem\footnote{For the cases where $y_h = 1/2$, for
odd number of spins, the zeros follow a more complicated pattern
(do not lie on a continuous curve apparently) and the numerical
uncertainties are larger.} to confirm the relation $\sigma=
(d-y_h)/y_h$. It may be appropriate to mention at this point that
there exists an analogous special point in the 2D Ising model for
which the natural variable is $\tilde{u}=u^2 = \exp -2\beta H$.
Namely, the point $\tilde{u}=-1 (\beta H_E = \pm \imath \pi/2) $
is also special, see \cite{ms}, there appears a change from
$\sigma =-1/6$ to $-1/9$ at $\tilde{u}=-1$. Henceforth, we
concentrate on the triple degeneracy case ($\sigma =-2/3$).

\section{Numerical results}

In this section we complement our analytic results in favor of a
new critical behavior with a numerical study. First of all, we
note that $\lbd_1=\lbd_2=\lbd_3=a_2/3$ holds if and only if :

\be a_1 \, a_2 - 9 \, a_0^2 \, = \, 0 \quad , \quad  3 \, a_1 -
a_2^2 \, = \, 0 \label{16} \ee

\no For the $S=1$ Ising model $(b=1=x)$ there is no real
temperature, not even in the antiferromagnetic case ($c\ge 1$), at
which conditions (\ref{16}) hold, similarly for the $q=3$ states
Potts model where $b=1/c^3  $ and $ \xt = u/c^3 $. However, for
the BEG model there are many solutions for the triple degeneracy
conditions (\ref{16}). In what follows we present numerical
results for two special cases.

\subsection{BEG model with $b=c \, (K=-J)$}

For the BEG model with $b=c \, (K=-J)$ the dependence of the cubic
equation on the magnetic field simplifies which allows us to find
the YLZ for a larger number of spins. In this case the conditions
(\ref{16}) imply the fine tuning

\be x=x(c)=(1+c^2)\sqrt{(1+c^2)/(27(1-c^2))} \label{xc1} \ee

\no and the position of the edge singularity: $A_E = \left\lbrack
2(4 c^2-5)/(1+c^2) \right\rbrack x(c)$. The singularity is on the
left-hand endpoint of an arc on the unit circle (fig. 1a) for $0
\le c \le \sqrt{2}/2 $ and on $\mathbb{R}_-\, $ for $ \sqrt{2}/2
\le c \le 1$, see figures 1a and 1b. We have checked that the YLES
on the right-hand endpoint of the arc is of the usual type $\sigma
= -1/2$.

In finding the zeros we have started from the expression for the
partition function of the BEG model obtained by a diagrammatic
expansion similar to what has been done in \cite{almeida} for the
Blume-Capel model. Following the same steps of \cite{almeida}, we
first define the generating function of partition functions of the
BEG model on non-connected rings and then we take its logarithm
which gives rise, as is well known in diagrammatic expansions, to
the generating function of only connected diagrams (usual rings).
Those steps lead us to $Z_N = - N\, c^{-N} \, \left\lbrace
\ln\left\lbrack 1 - a_2 \, g + a_1 \, g^2 - a_0 g^3 \right\rbrack
\right\rbrace_{g^N}$. The notation $\lbrace f(g) \rbrace_{g^N} $
indicates the  power $g^N$ in the Taylor expansion of $f(g)$ about
$g=0$. The constant $g$ plays the role of a coupling constant in a
diagrammatic expansion {\it a la} Feynman (perturbation theory).
 The above expression for $Z_N$ is equivalent to
(\ref{zn}) for arbitrary number of spins. In particular, the
reader can easily check it for low number of spins by using the
relationships between eigenvalues and coefficients of the cubic
equation given in the text between formulae (\ref{6}) and
(\ref{11a}). The new expression turns out to be more efficient
than (\ref{zn}) from a computational point of view. In practice,
we have used rings with $N_a = 210 + 10 (a-1)$ spins where $1\le
a\le 10 $. The zeros have been found with the help of Mathematica
software. They agree up to the first 30 digits with few cases
where they are known exactly as, for instance, $\xt =c(3-c^2)^2/2$
and $b= c(3-c^2)/2 $ where, for even number of spins, the Yang-Lee
zeros are exactly located at $A_k = -2 - 2\, i (1-c^2)\sin
\left(2\pi (1+3k)/3N\right)$ , $k=0,\cdots , N-1$.

According to the discussion at the end of the previous section, we
expect $y_h = 3/2$ at $c=\sqrt{2}/2$ where $A_E = -2$, and $y_h=3$
elsewhere. This is approximately confirmed in table 1. The
estimates $y_h^{(1)}$ and $y_h^{(2)}$ refer respectively to di-log
fits of (\ref{un}) and (\ref{rhon}) with $L=N$ and $d=1$. We take
the logarithm of the absolute value of the imaginary (real) part
of (\ref{un}) for $0\le c\le \sqrt{2}/2$ ($\sqrt{2}/2 < c \le 1$)
and neglect corrections to scaling in (\ref{un}) and (\ref{rhon}).
For both fits we have calculated $\chi^2$ and the deviation from
the ideal ($\pm 1$) Pearson coefficient, but since they have the
same temperature dependence we only display $\chi^2$ in our
tables. For $0.1 \le c \le 0.5 $ the results are quite homogenous
but the proximity of the point $\sqrt{2}/2 \approx 0.707 $ leads
to a crossover behavior at $c=0.7$ with an increase of 4 orders of
magnitude in $\chi^2$.

Alternatively, from (\ref{rhon}) with $d=1$ and $L= N$, neglecting
corrections, we obtain\footnote{A similar formula can be deduced
from ($\ref{un}$) but it gives  results even closer to $y_h=3$.}:

\be y_h^{(\rho)} (N_a) = 1 + \left\lbrack
\ln\frac{N_{a+1}}{N_a}\right\rbrack^{-1}\ln \left\lbrack
\frac{\rho(N_{a+1})}{\rho(N_{a})}\right\rbrack  \qquad , \qquad
a=1,2,\cdots,9  \label{20} \ee

\no Some results are shown in table 2 altogether with the $N\to\infty$ Bulirsch-Stoer (BST) \cite{bst,schutz}
extrapolation. We remark that in the BST approach there is a free parameter $w$ which depends on how we approach
the $N\to \infty$ limit, namely, $y_h(N) = y_h(\infty) + \frac{A_1}{N^w} + \frac{A_2}{N^{2w}} + \cdots $. The
BST algorithm approximates the original function $y_h(N)$ by a sequence of ratios of polynomials with a faster
convergence than the original sequence $y_h(N_a)$. Following \cite{schutz} we can determine $w$
 in a such way that the estimated uncertainty of the
 extrapolation is minimized. By defining the uncertainty as twice
the difference
  between the two approximants just before the final extrapolated
   result, see \cite{schutz}, we have
tested $w$ in the range $0.5 \le w \le 2.5$ and concluded that $w=1$ is the best choice. All BST results in this
work have been calculated at $w=1$. For $0.1 \le c \le 0.6$ the extrapolations are remarkably close to $y_h=3$.
The huge uncertainty at $c=0.7$ (last row in the third column in table 2) signalizes again the crossover to
$y_h=3/2$. Our numerical results confirm both finite size scaling relations (\ref{un}) and (\ref{rhon}). See
fig.2 for a plot regarding the relation (\ref{rhon}).

Before we go to the next case we can, without much numerical
effort, check that the triple degeneracy conditions lead in fact
to new results. When we choose $b=c$ the equation (\ref{fphi})
becomes a cubic equation for $A(\varphi )$. By further fixing $x =
x (c)$ according to (\ref{xc1}) and $c=0.4$, for sake of
comparison with the zeros in fig.1, we find three solutions
$A_i(\varphi ) \, , \, i=1,2,3 \, $ which are still complicated,
but expanding about $\varphi=0$ we derive:

\bea A_1(\varphi ) \, &=& \, \frac{829}{900}\sqrt{\frac{29}7} -
\frac{9 \sqrt{203}}{400} \varphi^2 + \frac{\sqrt{203}}{1440}
\varphi^4 + {\cal{O}}\left(\varphi^6\right) \label{A1} \\
A_2(\varphi ) \, &=& \, -\frac{218}{225}\sqrt{\frac{29}7} +
\frac{1}{75} \sqrt{\frac{203}{3}} \varphi^3 -
\frac{\sqrt{203}}{450}
\varphi^4 + {\cal{O}}\left(\varphi^5\right) \label{A2} \\
A_3(\varphi ) \, &=& \, A_2(-\varphi ) \label{A3} \eea

\no Now we see that the symmetry $\varphi \to -\varphi $ of the
equation (\ref{fphi}) mentioned in \cite{ananikian} does not
necessarily lead to even solutions. The operation  $\varphi \to
-\varphi $ exchange the second with the third solution. The first
solution $A_1(\varphi)$ is the usual Yang-Lee edge singularity and
gives $\sigma = -1/2$ while the second one leads to $\sigma =
-2/3$. The second solution corresponds to the new edge singularity
appearing on the left side of the arcs in fig.1a. The third
solution is not a true singularity. We can, even before the
expansion about $\varphi=0$, substitute $A_i(\varphi)$ in the
eigenvalue solutions of the cubic equation (\ref{cubic}) and check
numerically that the magnitude of two eigenvalues degenerates for
all $A_i(\varphi)$. However, for $A_3(\varphi)$ such magnitude is
smaller than the real (non-degenerate in magnitude) eigenvalue and
so it does not satisfy (\ref{5b}).

\subsection{Blume-Capel model ($b=1$)}

For the Blume-Capel model $K=0 \, (b=1)$ the triple degeneracy
conditions (\ref{16}) lead to complicated formulas for both the
fine tuning function $\xt = \xt (c) $ and the position of the new
edge $ A_E $ :

\be \xt (c) = \left\lbrace 1-8 c+10 c^2 + 10 c^3 - 9 c^4 +
3(1-c^2)\left\lbrack (P+Q)^{1/3} \, + R\, (P+Q)^{-1/3}\
\right\rbrack \right\rbrace^{1/2}\label{xc2} \ee

\be A_E = \frac{\xt (c) \left\lbrack (1+c)(5-15 c+ 5 c^2 + 3 c^3)
+ \xt^2 (c)\right\rbrack}{(1+c)(1+c^2) \, + \, \xt^2(c)(2-3c)}
\label{ae2} \ee

\no Where $P=\sqrt{c^3(1+c-c^2)(1-3c-c^2)^2} \, $  , $\,
Q=c^2(5-36 c + 95 c^2-90 c^3 + 27 c^4)$ and $R = c(9 c^3-20 c^2 +
10 c -1)$. In the whole range of temperatures $0\le c \le 1$  the
function $\xt (c)$ given in (\ref{xc2}) runs from +1 to +2, in
fact we have approximately\footnote{Although $\xt (c) \approx 1 +
c$ it is important that $\xt (c) \ne 1 + c$  since $\xt =
(1-c^2)b^2/\vert b - c \vert = 1+ c$ for $b=1$ and, as we have
discussed after formula (\ref{14b}), at that point we may non
longer have $dA/d\varphi =0$ at $\varphi=0$. In fact, we have
confirmed numerically that $y_h\approx 1$ at this point. It
corresponds to a bifurcation point where one curve becomes two
disjoint arcs of zeros, see fig.(2c) of \cite{almeida}} $\xt (c)
\approx 1 + c$.
 The position of the zeros
 $A_E$ ranges from  -2 to +2 which guarantees, see (\ref{A}),
  that the new edge is always on
the unit circle on the complex u-plane. However, the finite size
numerical calculations reveal that the zeros on the unit circle
correspond to a small fraction of whole set of zeros, see fig.3 .
Those zeros tend to form a short arc whose right-hand endpoint is
a usual YLES with $\sigma = -1/2$ as we have checked. The unusual
behavior appears on the left-hand endpoint, whose results are
displayed in tables 3 and 4. There is no edge singularity for the
other zeros outside the unit circle. For technical numerical
reasons we have used $ N_a = 96 + (a-1)6$ spins with $ 1 \le a \le
10$ and $0.3 \le c \le 0.9$. Although the finite size results move
away from $y_h=3$ as $T\to 0$ (table 3), the BST extrapolations
(table 4) all tend to $y_h=3$ with increasing uncertainty as $T\to
0$. In table 4 we only display the limit cases $c=0.3$ and
$c=0.9$. Since we have noticed that the fraction of zeros on $S_1$
decreases as $T\to 0$, our interpretation is that we are closer to
a continuum distribution of zeros as the temperature increases and
the temperature dependence of $y_h^{(\rho)}(N)$ is a pure finite
size artefact. In conclusion, our numerical results are in good
agreement with our analytic predictions.

Analogously to what we have done at the end of last subsection, if
we now assume $b=1$, $\xt =\xt (c)$ as given in (\ref{xc2}) and
fix some specific value for the temperature, we show that the
expansions of the solutions of the equation (\ref{fphi}) around
$\varphi=0$ give both $\sigma = -1/2$ and $\sigma = -2/3$.

\section{Conclusion}

We have derived for the Blume-Capel and BEG models, analytically and numerically, a new value for the magnetic
scaling exponent $y_h = 3$ and edge exponent $\sigma = -2/3$. This is only possible due to a triple degeneracy
of the transfer matrix eigenvalues. We have checked that our di-log fits, see e.g. fig.2, are consistent with
the finite size scaling relations (\ref{un}) and (\ref{rhon}) which appear around the usual Yang-Lee edge
singularity. We stress that the new critical behavior found here is not in conflict with previous numerical and
analytic results in the literature. In particular, the proof of $\sigma = -1/2$ given in
\cite{fisher2,kurze,wang} assumes that in the gap of zeros close to the YLES all eigenvalues are real. In the
examples that we have analyzed so far we have verified that this is indeed the case for the usual YLES ($\sigma
= -1/2$), however only the largest eigenvalue is real in the gap close to the new singularity ($\sigma = -2/3$).
Besides, although all eigenvalues are real at vanishing magnetic field, as one can check from (\ref{cubic}) at
$A=2$, apparently nothing prevents the two eigenvalues with smallest magnitude of acquiring an imaginary part
before reaching the edge singularity as we increase the magnetic field through imaginary values. Consequently,
so far we have not been able to prove that the edge singularity closer to $\mathbb{R}_+$ is always of the usual
type $(\sigma = -1/2)$, or alternatively, that the triple degeneracy condition could be satisfied also by an
edge singularity close to the positive real axis. Concerning the recent proof of $\sigma = -1/2$ given in
\cite{ananikian}, it supposes that $\left( d^2 u/ d\varphi^2\right)_{\varphi =0}\ne 0$ which is not the case in
the examples presented here where $\lbd_1=\lbd_2=\lbd_3 \, \ne \, \xt (b-c)/b$.

Finally, it is important to remark that instead of imposing the
triple degeneracy conditions (\ref{16}) and finding rather
peculiar fine tuning functions (\ref{xc1}) and (\ref{xc2}) we
could turn the argument around and claim that for some given
values of the couplings $b$ and $x$ we can, in many cases, find a
specific real temperature $c$ such that conditions (\ref{16}) are
satisfied and the usual result $\sigma = -1/2$ is replaced by
$\sigma = -2/3$. Since the number of transfer matrix eigenvalues
is infinite for $d>1$, even for the $S=1/2$ Ising model, we might
speculate that the degeneracy of more than two largest eigenvalues
could play a role also in higher dimensional spin models and lead
to a change in the exponent $\sigma$ at specific temperatures in
$d>1$.

\section{Acknowledgments}

D.D. is partially supported by CNPq and F.L.S. has been supported
by CAPES. D.D. has benefited from discussions with A.S. Castro,
R.V. de Moraes, A. de Souza Dutra and M. Hott. A discussion with
N.A. Alves on the BST extrapolation method is gratefully
acknowledged. Special thanks go to Luca B. Dalmazi for some
computer runs.

\vfill\eject

\begin{figure}
\begin{center}
\epsfig{figure=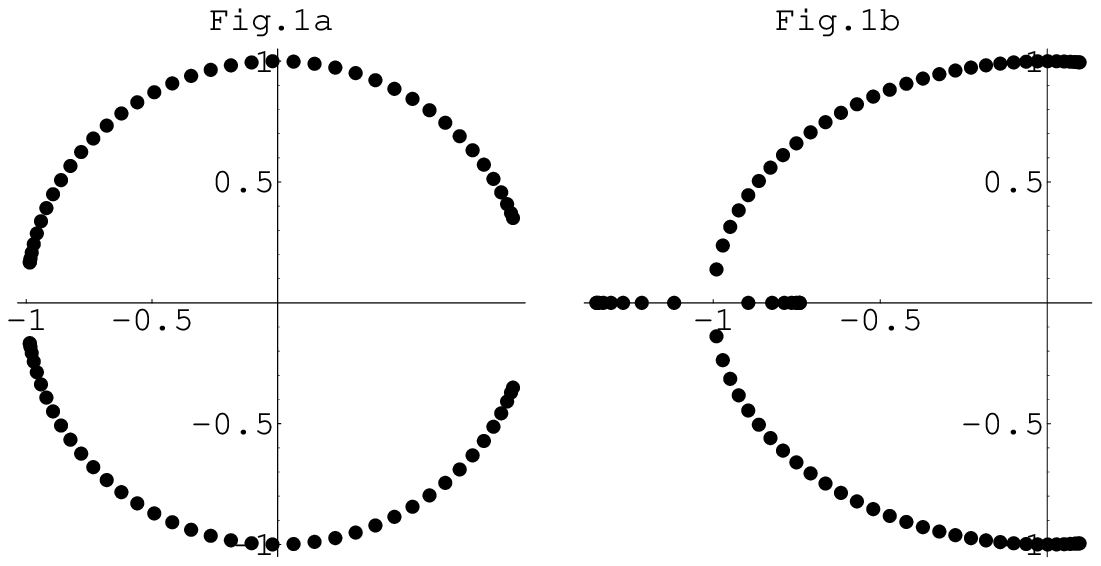,width=120mm} \caption{Yang-Lee zeros for
the BEG model with $b=c$  on the complex u-plane $(u=\exp^{-\beta
H})$ at the triple degeneracy condition, see (\ref{xc1}), and
temperatures $c=0.4$ (Fig.1a), $c=0.9$ (Fig.1b). In both figures
we have $100$ zeros. The closest zero to the new edge corresponds
to $u_1^{\pm}\approx -0.98602 \pm 0.16660 \, i$ in Fig.1a and
$u_1^+\approx-0.74063 \, , \, u_1^-\approx-1.35021$ in Fig.1b
}\label{fig2}
\end{center}
\end{figure}

\begin{figure}
\begin{center}
\epsfig{figure=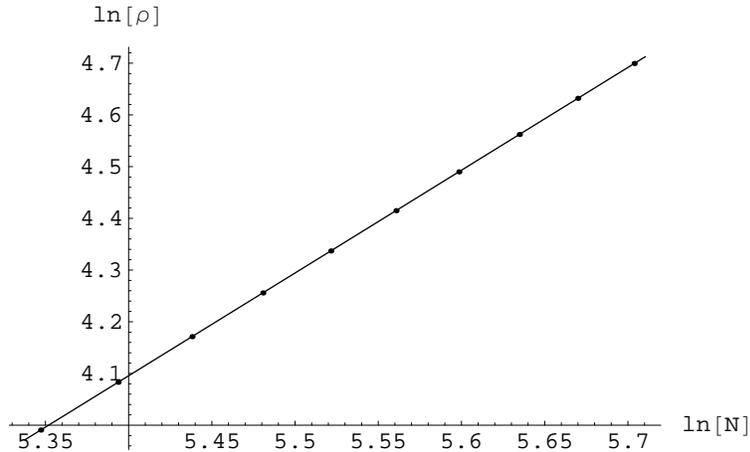,width=100mm} \caption{Di-log fit for the
density of zeros around the new edge singularity for the $\Delta =
-J \, (b=c) $ BEG model. We have used
 $210 \le N \le 300$ spins at temperature $c=0.6$. The solid line
is $\ln \rho = 1.98724 \left(\ln N\right) - 6.63547 $.
}\label{fig1}
\end{center}
\end{figure}

\begin{table}
\begin{center}
\begin{tabular}{|c|c|c|c|c|}
 \hline
 $c$ & $ y_h^{(1)}$ &  $10^{9} \chi^2_{(1)}$ &   $ y_h^{(2)}$ &  $10^{8} \chi^2_{(2)}$ \\
\hline
 0.1-0.5 & 2.997 & 3.5 & 2.989  & 4.0  \\
 0.6 & 2.997 & 3.5 & 2.987  & 8.3 \\
0.7 & 2.880 & 3.2 $\times 10^4$  & 2.094  & 2.4$\times 10^4$ \\
$\sqrt{2}/2$ & 1.498 & 0.9 & 1.494 & 1.3 \\
0.8 & 3.000 & 3.3 & 2.989  & 2.3 \\
0.9 & 2.997 & 3.5 & 2.989  & 3.8  \\
\hline
\end{tabular}
\end{center}
\caption{Data from di-log fits of formulae (\ref{un}) and
(\ref{rhon}) with $L=N$ and $d=1$ for the $\Delta = -J \, (b=c)$
BEG model with $210 \le N \le 300$ spins.}\label{tab1}
\end{table}

\begin{table}
\begin{center}
\begin{tabular}{|c|c|c|}
\hline
 $N_a$ & $ y_h^{(\rho)}(c=0.6)$ &    $ y_h^{(\rho)} (c=0.7)$ \\
\hline
210 & 2.98450354337 & 1.96268002086 \\
220 & 2.98540244520 & 1.99853987788  \\
230 & 2.98619971860 & 2.03403890400 \\
240 & 2.98691163425 & 2.06907936446 \\
250 & 2.98755121168 & 2.10357440914  \\
260 & 2.98812897995 & 2.13744702199\\
270 & 2.98865353760 & 2.17062929861\\
280 & 2.98913197003 & 2.20306194085 \\
290 & 2.98957016460 & 2.23469387995\\
$\infty$ & 3.000009(6) & 3(8) \\
\hline
\end{tabular}
\end{center}
\caption{Finite size results, obtained from (\ref{20}), for the
magnetic scaling exponent $y_h$ for the ($b=c$) BEG model with
$210 \le N \le 300$ spins at temperatures $c=0.6$ and
$c=0.7$.}\label{tab2}
\end{table}

\begin{figure}
\begin{center}
\epsfig{figure=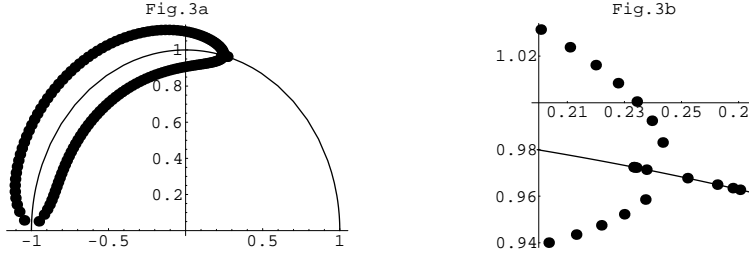,width=120mm} \caption{Half (Im$(u)>0$) of
the Yang-Lee zeros for the Blume-Capel model $(b=1)$ on the
complex u-plane $(u=\exp^{-\beta H})$
 at the triple degeneracy condition, see (\ref{xc2}), temperature $c=0.5$ and $N=150$ spins. The solid line in both
figures is part of the unit circle. The closest zero to the new
edge corresponds to $u_1^{+} \approx 0.2332 + 0.9734\, i$ as shown
in detail in Fig.3b. }\label{fig3}
\end{center}
\end{figure}

\begin{table}
\begin{center}
\begin{tabular}{|c|c|c|c|c|}
\hline
 $c$ & $ y_h^{(1)}$ &  $ 10^5 \chi^2_{(1)}$ &   $ y_h^{(2)}$ &  $10^{3} \chi^2_{(2)}$ \\
\hline
 0.3 & 3.200 & 5.03 & 4.612  & 3.10 \\
 0.4 & 3.148 & 2.52 & 3.795  & 2.10\\
0.5 & 3.119 & 1.51  & 3.561  & 0.72 \\
0.6 & 3.100 & 1.00 & 3.442 & 0.36 \\
0.7 & 3.088 & 0.76 & 3.370  & 0.22\\
0.8 & 3.078 & 0.58 & 3.320  & 0.15 \\
0.9 & 3.071 & 0.48 & 3.285  & 0.11  \\
\hline
\end{tabular}
\end{center}
\caption{Data from di-log fits of formulae (\ref{un}) and
(\ref{rhon}), with $L=N$ and $d=1$, for the Blume-Capel model with
$96 \le N \le 150$ spins and temperatures $0.3 \le c \le 0.9$. We
have used $ \vert \Re e (u_1^+(N) - u_1(\infty)) \vert $ in the
di-log fit of (\ref{un}). }\label{tab3}
\end{table}

\begin{table}
\begin{center}
\begin{tabular}{|c|c|c|}
\hline
 $N_a$ & $ y_h^{(\rho)}(c=0.3)$ &    $ y_h^{(\rho)} (c=0.9)$ \\
\hline
 96 & 6.65138835174 & 3.36602639165 \\
102 & 5.40615603620 & 3.33723696161 \\
108 & 4.85302919346 & 3.31271060187 \\
114 & 4.52543971437 & 3.29155556642 \\
120 & 4.30452579272 & 3.27311462040  \\
126 & 4.14377246532 & 3.25689223008\\
132 & 4.02074088023 & 3.24250742353 \\
138 & 3.92311802198 & 3.22966230906 \\
144 & 3.84352213555 & 3.21812047991 \\
$\infty$ & 3.001(1) & 3.00000(8) \\
\hline
\end{tabular}
\end{center}
\caption{Finite size results, obtained from (\ref{20}), for the
magnetic scaling exponent $y_h$ for the Blume-Capel model ($b=1$)
with $96 \le N \le 150$ spins at temperatures $c=0.3$ and
$c=0.9$.}\label{tab4}
\end{table}

\end{document}